\documentstyle[11pt,paspconf,epsf]{article}

\begin{document}

\title
{
{\it Ab initio} galaxy formation
}

\author
{
C. M. Baugh, A. J. Benson, S. Cole, C. S. Frenk, C. G. Lacey 
}
\affil{Dept. of Physics, Durham University, South Road, DH1 3LE, England}

\begin{abstract}
The formation and evolution of galaxies can be followed in the 
context of cosmological structure formation using the technique 
of semi-analytic modelling. 
We give a brief outline of the features incorporated 
into the semi-analytic model of Cole etal (1999). 
We present two examples of model predictions that can be tested 
using photometric redshift techniques. 
The first prediction, of the star formation history of the universe, 
has already been shown to be in broad argeement with the observational 
estimates.
The second prediction, of the evolution of galaxy clustering 
with redshift, will be addressed with some of the forthcoming deep,  
multi-filter imaging surveys discussed at this meeting.
\end{abstract}


\section{Introduction}

The main ideas of hierarchical galaxy formation were set out more than 
twenty years ago by White \& Rees (1978). 
These authors proposed that galaxy formation is a two stage process. 
In the first stage, dark matter haloes form by the dissipationless 
accretion of smaller units and through mergers. 
The second stage consists of the dissipative condensation of baryons to the 
centre of dark matter haloes.
After the first stars form from this gas, feedback effects can play 
an important role, regulating star formation and thus 
controlling the efficiency of galaxy formation in dark matter 
haloes of different mass.

The past decade has seen the development of semi-analytic models 
to study the {\it ab initio} formation and evolution of galaxies, 
within the framework of the growth of structure in the dark 
matter (e.g. Kauffmann etal 1993, Cole etal 1994, Somerville \& 
Primack 1998).
The dissipationless physics in these models is well understood and 
has been explored extensively using N-body simulations of gravitational 
instability (for recent illustrative examples 
see Jenkins etal 1998 and Ghigna etal 1999).
The dissipative processes are, however, not  
at all well understood. 
The idea that hot gas would cool radiatively to make galaxies 
was also proposed in the late 1970s, but numerical simulations of 
this process with sufficient resolution to identify ``galaxies'' 
within cosmological volumes are only now becoming possible (Pearce etal 1999).
In the semi-analytic models, the processes of gas cooling, star formation, 
the attendant feedback and galaxy mergers are described by a set of 
simple, physically motivated rules.
The values of the required parameters are set by comparing the model results 
with properties of the local galaxy population, such as the field 
galaxy luminosity function or the Tully-Fisher 
relation (see Somerville \& Primack for a discussion 
of the various models), to produce a fully specified model with strong 
predictive capabilities.

\section{{\it Ab initio } galaxy formation}

The semi-analytic model described by Cole etal (1999) is a development 
of the one described in earlier papers by the Durham group 
(e.g. Cole etal 1994, Baugh, Cole \& Frenk 1996). 
In addition to the use of improved algorithms, for example in the 
Monte-Carlo generation of dark matter halo merger histories, 
a broader range of physical processes is now modelled, 
greatly expanding the number of galaxy properties that 
we can predict.

The size of a galactic disk is determined by the conservation of the angular 
momentum of cooling gas. The gas is assumed to have the same specific 
angular momentum profile as the dark matter halo, whose spin originates from 
tidal torques that act during its formation.
The size of a bulge is computed by conserving energy when two 
fragments merge to form the bulge and applying the virial theorem.
The condensation of baryonic material at the centre of a dark matter 
halo alters the structure of the halo, causing a contraction.
The resulting size of the disk and bulge components depends upon the 
self-gravity of the baryons and the gravity of the modified dark 
matter halo.

The chemical enrichment of the baryons is also followed in the 
new model. Episodes of star formation result in the production 
of metals that can be transferred to the reservoirs of hot and 
cold gas within each dark matter halo. Feedback from star formation 
can reheat some of the cold gas, thus providing a further 
channel through which to transfer metals to the hot gas component.
The calculation of the scale length of the galactic disk and of the 
metallicity of the cold gas allows us to obtain an optical depth for the 
disk, and hence to calculate the extinction of starlight by dust.

\begin{figure}
\plotfiddle{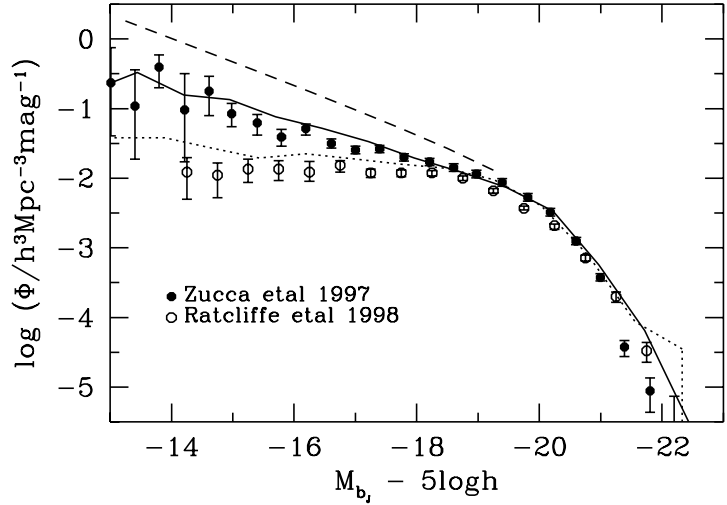}{4.6truecm}{0.0}{60}{60}{-180}{-160}
\caption
{
The points show two recent determinations of the local field galaxy 
luminosity function. The solid line shows the luminosity function 
in the fiducial model of Cole etal (1999); the dotted line shows the 
luminosity function obtained when feedback is parameterised as a 
much stronger function of circular velocity, as used by Cole etal. (1994).
The dashed line indicates the slope of the dark matter halo mass function. 
} 
\label{fig:lf}
\end{figure}

The main property of the local galaxy population that 
we use to constrain our models is the field galaxy luminosity function. 
Figure \ref{fig:lf} shows two recent determinations of the 
luminosity function in the $b_{J}$-band. These surveys are taken  
from the same parent photometric catalogue, but extend to different 
apparent magnitude limits. 
The solid line shows the luminosity function of the fiducial model of
Cole etal (1999), 
which is in very good agreement with the 
measurement by Zucca etal (1997) over a dynamic range of $\sim 10^{4}$ 
in luminosity and $\sim 10^{5}$ in space density. 
The dotted line shows a model in which the feedback arising from star 
formation is assumed to be a much stronger function of 
circular velocity, as in the model of  
Cole etal (1994). Star formation in low circular velocity objects 
is extremely inefficient in this case, and a reasonable match is obtained 
to the flat faint end slope of the luminosity function found by Ratcliffe 
etal (1998). 
The dashed line shows the slope of the dark matter halo mass function; 
this slope, $\alpha \sim -1.8$, would be obtained for the luminosity function 
in the absence of feedback and if galaxies merged on the same timescale 
as their parent dark matter haloes.

\section{The star formation history of the universe}

The use of the redshifted Lyman-break spectral feature to 
isolate candidate high redshift galaxies has proven to 
be a remarkably successful and efficient way of constructing 
large samples of galaxies at significant lookback times 
(Steidel etal 1996 and references therein). 
The application of this photometric technique to deep images obtained from 
the ground and to the {\it Hubble Deep Fields} has allowed the star 
formation history of the universe to be constructed 
(e.g. Madau etal 1996, Steidel etal 1999).

\begin{figure}
\plotfiddle{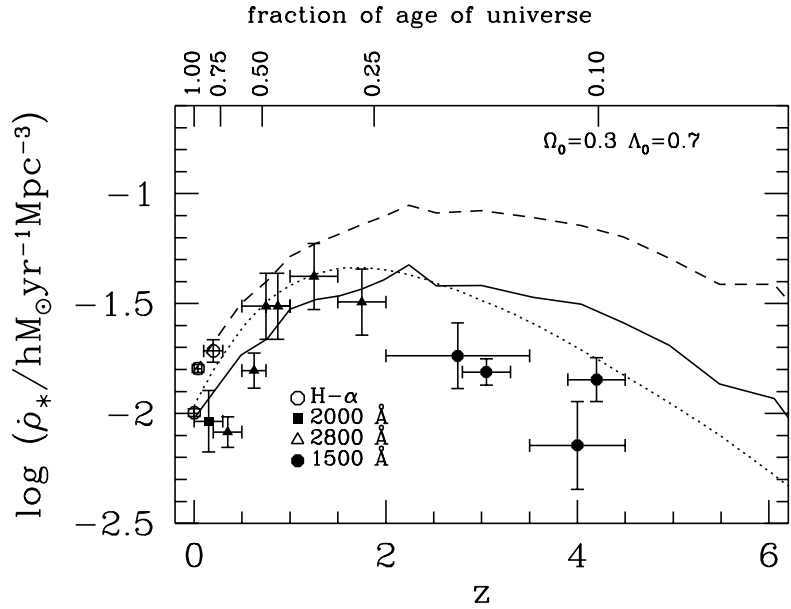}{4.6truecm}{0.0}{60}{60}{-180}{-170}
\caption
{
The star formation history of the universe. The dotted line 
shows the prediction of Cole etal (1994). The dashed line shows 
the star formation rate density in the model of 
Cole etal (1999), inferred from the $1500\AA$ luminosity density. 
The solid line shows the effect on this estimate of obscuration of 
UV starlight by dust in the model.
The points show a subset of recent observational determinations of the 
star formation density, using various indicators of the star formation 
rate as indicated by the key. 
} 
\label{fig:sfrv}
\end{figure}

Figure \ref{fig:sfrv} shows a subset of the currently available 
observational estimates of the star formation rate per unit volume, 
as inferred from the flux density at various wavelengths, 
indicated by the key in the Figure.
The dotted line shows the star formation history predicted by the model 
of Cole etal (1994), which predates the oldest data points on the 
Figure by two years (see Figure 16 of Baugh etal 1998).

The dashed line shows the star formation history in the fiducial 
model of Cole etal (1999), estimated from the rest-frame luminosity 
density at $1500\AA$. 
The solid line shows how this estimate is changed when obscuration 
by dust is taken into account; this line is the one that should 
be compared to the data points, which have not been 
corrected for the effects of dust. 
The star formation rate per unit volume in the model is reduced by 
a factor of $\approx 2.5$ at $z=4$ as a result of dust extinction 
of the $1500\AA$ light. 

The small differences between the dotted line (Cole etal 1994) and 
the solid line are mainly due to the different parameterisations of 
the strength of feedback as a function of circular velocity used 
in the two models. 
The particular choice of feedback model is motivated by the 
attempt to reproduce the faint end of the local luminosity function 
(see Figure \ref{fig:lf} and the discussion in the previous Section).
The weaker feedback employed by Cole etal (1999) 
results in a somewhat broader peak in the global star formation 
rate around $z \approx 2-3$, compared with the dotted line, which 
peaks at $z \approx 1.5-2$. 
However, when expressed in terms of lookback time (as shown by the scale 
at the top of the Figure), which is the appropriate variable for 
computing the integrated stellar mass, this difference 
is in fact quite small.

\section{The evolution of galaxy clustering}

Two approaches have been used to obtain predictions for the 
clustering of galaxies from semi-analytic models. 
In the first approach, a bias parameter is computed for dark matter 
halos to relate the amplitude of the correlation function of the haloes 
to that of the underlying dark matter distribution (Mo \& White 1996).
The semi-analytic model is then used to populate dark matter haloes 
with galaxies, and a selection criterion, such as an apparent 
magnitude limit, is applied to the model galaxies. 
The bias parameter for the galaxies is obtained by summing 
the bias of each halo, weighted by its abundance and the 
number of galaxies it contains that satisfy the selection 
criterion (e.g. Baugh etal 1999). 
This approach was used by Baugh etal (1998) to predict successfully  
that Lyman break galaxies have a correlation length similar to 
that of bright galaxies at the present day, and much larger than that 
of the dark matter in any viable model at $z \sim 3 $.

Another  method is to use the semi-analytic model 
to populate dark matter haloes in a N-body simulation of 
hierarchical clustering with galaxies (Kauffman etal 1997, 1999; 
Governato etal 1998, Benson etal 1999). 
This approach has the advantage of being able to probe 
galaxy clustering down to small scales and to include the distortion 
of the clustering pattern that arises from the peculiar motions 
of galaxies.
Benson etal (1999) find that the clustering of galaxies in the 
fiducial model of Cole etal (1999) is in very good agreement with 
the two point correlation function measured for APM galaxies. 
This is a remarkable result because the measured correlation function is a 
power law over more than a decade in separation, whereas the correlation 
function of the dark matter has two inflection points over the same 
range of scales. Furthermore, on scales around $1h^{-1}$Mpc, the amplitude 
of the dark matter correlation function is higher than that measured for 
galaxies, implying that galaxies are $anti$-biased or less clustered 
compared to the dark matter. The level of small scale clustering 
depends sensitively on how dark matter haloes are populated by 
galaxies, and thus upon the details of the galaxy formation process.

\begin{figure}
\plotfiddle{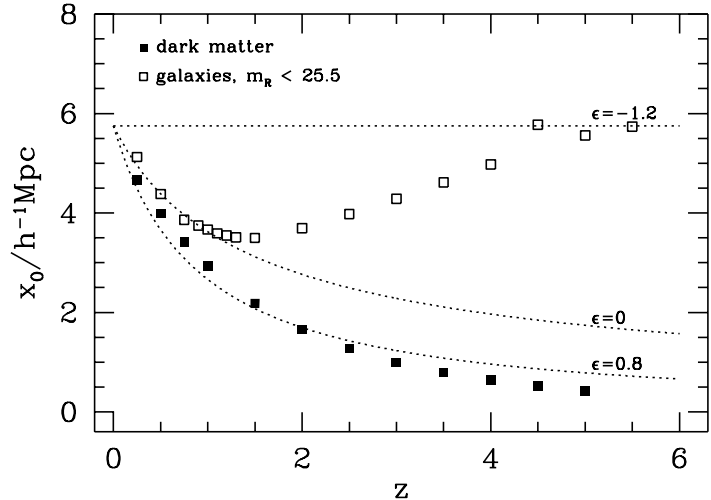}{4.5truecm}{0.0}{60}{60}{-190}{-170}
\caption
{
The evolution of the correlation length of clustering, 
defined as the {\it comoving} length scale for which the two 
point correlation function is unity: $\xi(x_{0}) = 1$ (Baugh etal 1999).
The filled squares show the evolution of the clustering length of dark 
matter in an $\Omega_{0}=0.3$, $\Lambda_{0}=0.7$ universe, with density 
fluctuations normalised to reproduce the observed local abundance 
of hot X-ray clusters.
The open squares show the evolution of the correlation length of galaxies 
in the model, selected to be brighter than $m_{R}=25.5$.
The dotted lines show the prediction of the $\epsilon$-model, for typical 
values of the $\epsilon$-parameter.
} 
\label{fig:xir}
\end{figure}

The first approach described above has been used to predict 
the evolution of the correlation length of galaxies, measured in comoving 
units, with redshift, as shown in Figure \ref{fig:xir} (Baugh etal 1999; 
similar results are found using the N-body technique, Kauffmann etal 1998). 
The clustering evolution of galaxies is markedly different 
to that displayed by the dark matter, and furthermore is not 
well described by the `$\epsilon$-model' commonly used to interpret 
such data. 
The correlation length of galaxies initially 
decreases to $z \approx 1$, and then starts to increase 
again at higher redshifts. 
The model predicts that galaxies that are bright enough to be seen 
at high redshift are hosted by the most massive dark matter haloes 
in place at these redshifts. Such haloes are biased 
tracers of the dark matter distribution.

Traditionally, the angular clustering of galaxies imaged in one 
band has been quantified in terms of the correlation amplitude 
measured at a fixed angular scale plotted as a function of 
the limiting apparent magnitude of the sample. 
Even at very faint magnitudes, the median redshift of 
the sample does {\it not} increase rapidly and in addition, 
the clustering signal is diluted by projection effects.
A powerful observational technique that is now being 
pursued by many groups is to take deep images in many different filters 
and to measure the angular clustering of galaxies selected to lie 
within a limited baseline in photometric redshift. 
This method has been applied to study the evolution of galaxy 
clustering to redshift $z \sim 1$ (Connolly etal 1998) 
and beyond using the Hubble Deep Field (e.g. Magliocchetti \& Maddox 1998). 
At present, the fields studied are small and the results may be 
subject to sample variance.
The detection of the dip in the galaxy correlation length  
shown in Figure 3 would provide evidence for hierarchical 
galaxy formation, but it may tell us more about the details of the 
process of galaxy formation than about the underlying cosmology. 
In any event, the semi-analytic models described here will have an 
important part to play in the interpretation of such results.

\acknowledgments
CMB would like to thank the organisers for an enjoyable and informative 
meeting and for providing assistance to attend the workshop.

\end{document}